\documentclass[5p]{elsarticle}
\biboptions{numbers,sort&compress}

\usepackage{lineno,color,slashed,amsmath,amssymb}
\modulolinenumbers[5]

\journal{Phys.\ Lett.\ B.}

\newcommand{\nn}{\nonumber}
\newcommand{\go}{\tilde g}
\newcommand{\neuone}{\tilde\chi^0_{1}}
\newcommand{\neu}{\tilde\chi^0_{1,2}}
\newcommand{\neui}{\tilde\chi^0_{i}}
\newcommand{\tg}{\tilde G}
\newcommand{\pg}{\tilde G'}

\newcommand{\MET}{\slashed{E}_T}
\newcommand{\be}{\begin{equation}}
\newcommand{\ee}{\end{equation}}
\newcommand{\bea}{\begin{eqnarray}}
\newcommand{\eea}{\end{eqnarray}}

\begin{document}

\begin{frontmatter}

\title{$Z$-peaked excess in goldstini scenarios}

\author[tokyo]{Seng Pei Liew}
\author[vub]{Alberto Mariotti}
\author[vub]{Kentarou Mawatari}
\ead{kentarou.mawatari@vub.ac.be}
\author[kcl]{Kazuki Sakurai}
\author[vub]{Matthias Vereecken}

\address[tokyo]{Department of Physics, University of Tokyo, 
 Bunkyo-ku, Tokyo 113-0033, Japan}
\address[vub]{Theoretische Natuurkunde and IIHE/ELEM, 
 Vrije Universiteit Brussel,
 and International Solvay Institutes,\\
 Pleinlaan 2, B-1050 Brussels, Belgium}
\address[kcl]{Department of Physics, King's College London, 
 London WC2R 2LS, UK}


\begin{abstract}
We study a possible explanation of a 3.0 $\sigma$ excess recently
 reported by the ATLAS Collaboration in events with $Z$-peaked
 same-flavour opposite-sign lepton pair, jets and large missing
 transverse momentum in the context of gauge-mediated SUSY breaking with
 more than one hidden sector, the so-called goldstini scenario. 
In a certain parameter space, the gluino two-body decay chain
 $\tilde g\to g\tilde\chi^0_{1,2}\to gZ\tilde G'$ becomes dominant,
 where $\tilde\chi^0_{1,2}$ and $\tilde G'$ are the Higgsino-like
 neutralino and the massive pseudo-goldstino, respectively, and gluino
 pair production can contribute to the signal. 
We find that a mass spectrum such as $m_{\tilde g}\sim 1000$~GeV,
 $m_{\tilde\chi^0_{1,2}}\sim 800$~GeV and $m_{\tilde G'}\sim 600$~GeV
 demonstrates the rate and the distributions of the excess, without
 conflicting with the stringent constraints from jets plus missing
 energy analyses and with the CMS constraint on the identical final
 state. 
\end{abstract}


\end{frontmatter}

\vspace*{-11.5cm}
\noindent
{\small UT-15-22, KCL-PH-TH/2015-29, LCTS/2015-21}
\vspace*{10cm}

\section{Introduction}

While the LHC Run-I achieved a great success with the discovery of the
Higgs boson, most of the attempts to find new physics failed, and
hence we are eagerly waiting for results from the LHC Run-II, which has
just started and entered a new energy frontier. 

However, although no significant signs for new physics at the 7 and
8~TeV run have been reported so far, we should still keep our eye on the
possible faintest imprint.
The ATLAS Collaboration has recently reported a search for supersymmetry
(SUSY) with the final state containing a pair of same-flavour opposite-sign
(SFOS) leptons, jets and large missing transverse momentum ($\MET$)
at a centre-of-mass energy of 8 TeV~\cite{Aad:2015wqa}.
With data for an integrated luminosity of 20.3~fb$^{-1}$, 
an intriguing excess of 29 lepton pairs peaked at the invariant mass of
the $Z$ boson (``on-$Z$'') is observed, while $10.6 \pm 3.2$ pairs are
expected from the Standard Model (SM) prediction.
The excess corresponds to a significance of $3.0~\sigma$.

Interpretation of the excess with some models is not a straightforward
task, as null results from other searches are placing considerably
stringent constraints on the viable model parameter space.
The ATLAS on-$Z$ signal region is motivated by search for a pair
production of gluino in gauge mediation scenarios in the minimal
supersymmetric standard model (MSSM), where gluino decays
into a neutralino, which subsequently decays into a $Z$ boson
accompanied by a very light gravitino.
The gravitino escapes detection, leading to missing momentum.
While relatively light gluinos with $m_{\go}<1.2$~TeV are
required~\cite{Barenboim:2015afa}, studies presented
in~\cite{Ellwanger:2015hva,Allanach:2015xga} have shown that explaining
the excess within this scenario is difficult. 
At large gluino--neutralino mass splitting, gluino decays with a
significant branching ratio into a top quark, leading to appreciably
strong bounds from the stop searches. 
Even in the compressed gluino--neutralino mass region, the process with
the hadronic decay mode of the energetic $Z$ boson is constrained by the
dedicated jets plus $\MET$ searches. 
Given that situation, several solutions have been proposed in various
models, e.g. NMSSM~\cite{Ellwanger:2015hva,Cao:2015ara}, MSSM with
a light sbottom or stop~\cite{Kobakhidze:2015dra},
pMSSM~\cite{Cahill-Rowley:2015cha}, split SUSY~\cite{Lu:2015wwa}, as
well as non-SUSY models~\cite{Vignaroli:2015ama,Dobrescu:2015asa}. 

In this paper, we investigate an alternative SUSY model, i.e. a model
with multiple hidden sectors, the so-called goldstini
model~\cite{Cheung:2010mc,Benakli:2007zza,Cheung:2010qf,Argurio:2011hs}, 
especially in the context of gauge mediated SUSY breaking.  
We consider a gluino ($\go$), the lightest and second lightest
neutralinos ($\neu$), and a pseudo-goldstino ($\pg$) in the spectrum as
a simplified model.  
As we verify below, in a certain parameter space, the gluino two-body
decay chain  
\begin{align}
 \go\to g+\neu\to g+Z+\pg
\label{godecay}
\end{align}
becomes dominant, and gluino pair production can contribute to the
signal. 
Jets from the gluino decay and the hadronic $Z$ decay are softened when
the mass spectrum is compressed due to the massive nature of the
pseudo-goldstino. 
We find that there is a viable parameter space even after the multijet
plus $\MET$ constraint~\cite{Aad:2014wea}, as well as the CMS constraint
on the identical final state~\cite{Khachatryan:2015lwa}, are taken into
account.
We also show that the two-body gluino decay in~\eqref{godecay} provides
a better fit to the data for the distributions with respect to the three-body gluino decay $\tilde g\to q\bar q\tilde\chi^0_{1,2}$. 

The paper is organized as follows:
In Sec.~\ref{sec:model}, we briefly review the model involving two SUSY
breaking sectors, and verify the gluino two-body decay chain
in~\eqref{godecay}.
In Sec.~\ref{sec:analyses}, we recast the ATLAS on-$Z$ analysis for our
benchmark scenarios to find a viable parameter space. 
The constraints coming from other searches are also discussed. 
Sec.~\ref{sec:summary} is devoted to our summary.
In~\ref{sec:neudecay} we give the explicit formulas for the neutralino
decays, while in~\ref{sec:validation} we present the validation of the
implementation for the analyses.

\section{Model}\label{sec:model}

The model we consider, in order to explain the ATLAS $Z$-peaked excess, 
is a scenario of gauge mediation with more than one secluded SUSY
breaking sector. 
Each SUSY breaking sector contains a massless goldstino, i.e.~the
goldstone fermion of spontaneous SUSY breaking. 
In a two-sector scenario there are then two massless goldstini.
Models of goldstini have been studied 
in~\cite{Cheung:2010mc,Benakli:2007zza,Cheung:2010qf,Craig:2010yf,Cheng:2010mw,Argurio:2011hs,Thaler:2011me,Cheung:2011jq,Argurio:2011gu,Cheung:2011jq,Liu:2013sx,Ferretti:2013wya,Hikasa:2014yra,Liu:2014lda}, 
and the description of goldstini in gauge mediation models and the
quantum corrections to the pseudo-goldstino mass have been investigated
in~\cite{Argurio:2011hs}. 
Indeed, when taking into account the mediation mechanism and the
interaction of the SUSY breaking sector with the MSSM, a linear
combination of the goldstini becomes the true goldstino (the
longitudinal component of the gravitino), while the orthogonal
combination is a pseudo-goldstino. 
Denoting with $f_a$ and $\tilde G_a$ the SUSY breaking scales and the
goldstini of the two sectors respectively, 
the true (massless) goldstino and the pseudo-goldstino are
\begin{align}
\label{gld_pgld}
 \tilde G=\frac{1}{F} \big(f_1 \tilde G_1 +f_2 \tilde G_2 \big), 
 \quad 
 \tilde G'=\frac{1}{F} \big(-f_2 \tilde G_1 +f_1 \tilde G_2 \big),
\end{align}
where $F=\sqrt{f_1^2+f_2^2}$ is the total SUSY breaking scale.
The true goldstino (or gravitino) mass is related to the SUSY breaking
scale and the Planck scale as $m_{\tg}\propto F/M_{\rm Pl}$, and hence
the low-energy SUSY breaking scenario as in gauge mediation leads to a
very light goldstino.
On the other hand, the mass of the pseudo-goldstino is not protected by
any symmetry and generically receives relevant quantum corrections,
proportional to the SUSY breaking terms~\cite{Argurio:2011hs}.
In our paper, we consider the pseudo-goldstino mass to be of the order
of a few hundreds of GeV.  

The interaction of each of the goldstini with the particles in the
visible sector are determined by the supercurrent, and hence the
couplings of the true goldstino and of the pseudo-goldstino with the
MSSM states are also fixed in terms of the soft SUSY breaking masses.
The noteworthy feature is that the pseudo-goldstino couplings are
generically different with respect to the goldstino couplings. 
Indeed they depend on the relative contribution to the soft masses from
each of the SUSY breaking sector, and they can be enhanced without
changing the overall soft masses.
This implies that the decay of SUSY particles can be dominantly into a
massive pseudo-goldstino plus a SM particle, drastically changing the
final state topology with respect to the usual decay into the true
goldstino~\cite{Argurio:2011gu,Cheung:2011jq,Liu:2013sx,Ferretti:2013wya,Hikasa:2014yra,Liu:2014lda}. 

\begin{figure}
\center
 \includegraphics[width=.45\columnwidth]{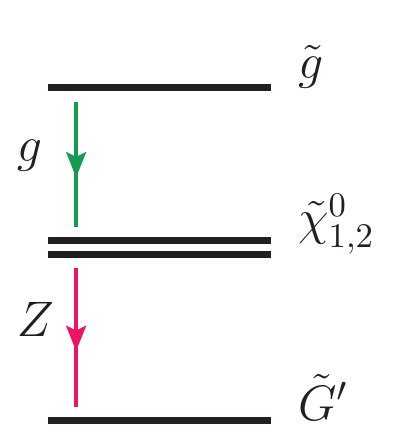}
\caption{\label{fig:spectra}
 Mass spectrum for our simplified goldstini model, with the relevant
 decay modes.}  
\end{figure}

In this paper we consider a simplified model including as low energy
degrees of freedom only a gluino, Higgsino-like neutralinos, a
pseudo-goldstino and a goldstino. 
The mass spectrum of the model is depicted in Fig.~\ref{fig:spectra}.
Note that the Higgsino fields include two almost degenerate neutral mass
eigenstates and a charged one. 
We also note that the true goldstino is in the bottom of the spectrum,
but is not shown in Fig.~\ref{fig:spectra} since it is irrelevant in our 
scenario.
In the following we study each decay step in detail to verify the gluino
decay chain in~\eqref{godecay}.

\paragraph{Gluino decay}

In the spectrum presented in Fig.~\ref{fig:spectra} the gluino can
potentially have several decay channels, depending on the mass splitting
between the gluino and the neutralinos. 

In case of the large splitting such as $(m_{\go}-m_{\neui})\gg m_t$, the
tree-level three-body decays into a pair of third-generation quarks and a
chargino/neutralino are dominant~\cite{Allanach:2015xga}. 

On the other hand, in the regime
\begin{align}
 \Delta m_{\go-\neu} \equiv m_{\tilde g}-m_{\neu}\leq 200 \mbox{ GeV},
\label{dmg}
\end{align}
the gluino decays predominantly into a gluon and a neutralino
via a (s)top loop.
The analytic expression for the $\go\to g\neui$ decay can be found 
in~\cite{Baer:1990sc,Gambino:2005eh}. 
We find that this result is robust as soon as the Bino ($M_B$) and Wino
($M_W$) masses are moderately larger than the Higgsino mass ($\mu$). 
We checked this decay pattern with SUSY-HIT~\cite{Djouadi:2006bz}, for
instance, by fixing $m_{\tilde g}=1000$~GeV, $\mu=800$~GeV, and
$\tan\beta=10$, we find 
$B(\go\to g\neu)>85\%$ as soon as $M_B$ and $M_W$ are larger than about
1.4~TeV, with squark masses of the order ${\cal O}(5)$~TeV.  

The gluino decays into a pseudo-goldstino or a goldstino with a gluon is 
also possible.
However, unless the gluino coupling to the pseudo-goldstino is largely
enhanced, these decay modes are always suppressed compared with the
decays into the MSSM states. 

In this paper, to fit the ATLAS excess, we consider the small mass
splitting in Eq.~\eqref{dmg}, where the only $\go\to g\neu$ decay is
relevant.

\paragraph{Neutralino decay}

The interaction lagrangian which is relevant for the neutralino decay 
into a pseudo-goldstino and a electroweak gauge boson or a Higgs boson 
is given 
by~\cite{Cheung:2010mc,Thaler:2011me,Argurio:2011gu,Liu:2013sx,Ferretti:2013wya}
\begin{align}
\mathcal{L}_{\pg}^{\rm int} &= 
  i\frac{\tilde y^i_{\gamma}}{2 \sqrt{2}F}\, 
  \tilde G' \sigma^{\mu} \bar \sigma^{\nu} \tilde \chi^0_{i} A_{\mu\nu} 
 +i\frac{\tilde y^{i}_{Z_T}}{2 \sqrt{2} F }\, 
  \tilde G'\sigma^{\mu} \bar \sigma^{\nu} \tilde \chi^0_{i} Z_{\mu\nu}  \nonumber \\
 &\quad+\frac{\tilde y^{i}_{Z_L} m_Z}{\sqrt{2} F }\, 
   \bar{\tilde{\chi}}^{0}_{i} \bar  \sigma^{\mu} \tilde G' Z_{\mu} 
 +\frac{\tilde y^{i}_{h}}{\sqrt{2} F }\,  \tilde \chi^0_{i} \tilde G' h
 +h.c.,
\label{pgld_couplings}
\end{align}
where $A_{\mu\nu}$ and $Z_{\mu\nu}$ are the field strengths of the
photon and the $Z$ boson, respectively, and $h$ is the lightest Higgs
boson in the decoupling limit. 
The goldstino lagrangian is the same, but with different coefficients,
$\tilde y\to y$.  
As mentioned above, the pseudo-goldstino couplings $\tilde y$ can be
larger than the goldstino couplings $y$, and in a simplified model
approach they can be considered as free
parameters~\cite{Argurio:2011gu,Liu:2013sx,Hikasa:2014yra}. 
Given a set of soft terms originating from the two SUSY breaking
sectors, one can compute these couplings with the formulas collected
in~\ref{sec:neudecay}. 

Although the neutralino decay can present a rich pattern, with six
competing decay modes as $(Z,h,\gamma)$ plus $(\pg,\tg)$, 
we are interested in a scenario where the neutralino predominantly
decays into a pseudo-goldstino and a $Z$ boson.
This scenario can be realized by enhancing the coupling parameters
$\tilde y$, especially $\tilde y_{Z_T}$ and
$\tilde y_{Z_L}$, and by assuming the Higgsino-like neutralino.
The decay formulas are reported in~\ref{sec:neudecay}, where we also 
provide the details of our illustrative benchmark point in the SUSY
breaking parameters determining the couplings $\tilde y$ and $y$;
we typically take $\mu\sim800$~GeV and $M_B=M_W\sim1.5$~TeV.
For the parameters we consider in this paper,
the mass splitting between the two neutralinos
is of the order of a few GeV.  
In this scenario, the second lightest neutralino $\tilde\chi_2^0$ decays
to the lightest neutralino $\tilde\chi^0_1$ with soft SM particle
emissions. 
Indeed we checked with SUSY-HIT~\cite{Djouadi:2006bz} and the formula
in~\ref{sec:neudecay} that its decay modes to the goldstino and to the
pseudo-goldstino for our benchmark point are negligible compared with
the $\tilde\chi^0_2\to\tilde\chi^0_1$ decays.
Hence in the following we assume 
$B(\tilde\chi^0_2\to\tilde\chi^0_1+{\rm undetectable\ SM\ particles})=100\%$.
The final state topology is then determined by the possible
$\tilde\chi^0_1$ decays, which we now investigate.

\begin{figure}
\center
\includegraphics[width=.88\columnwidth]{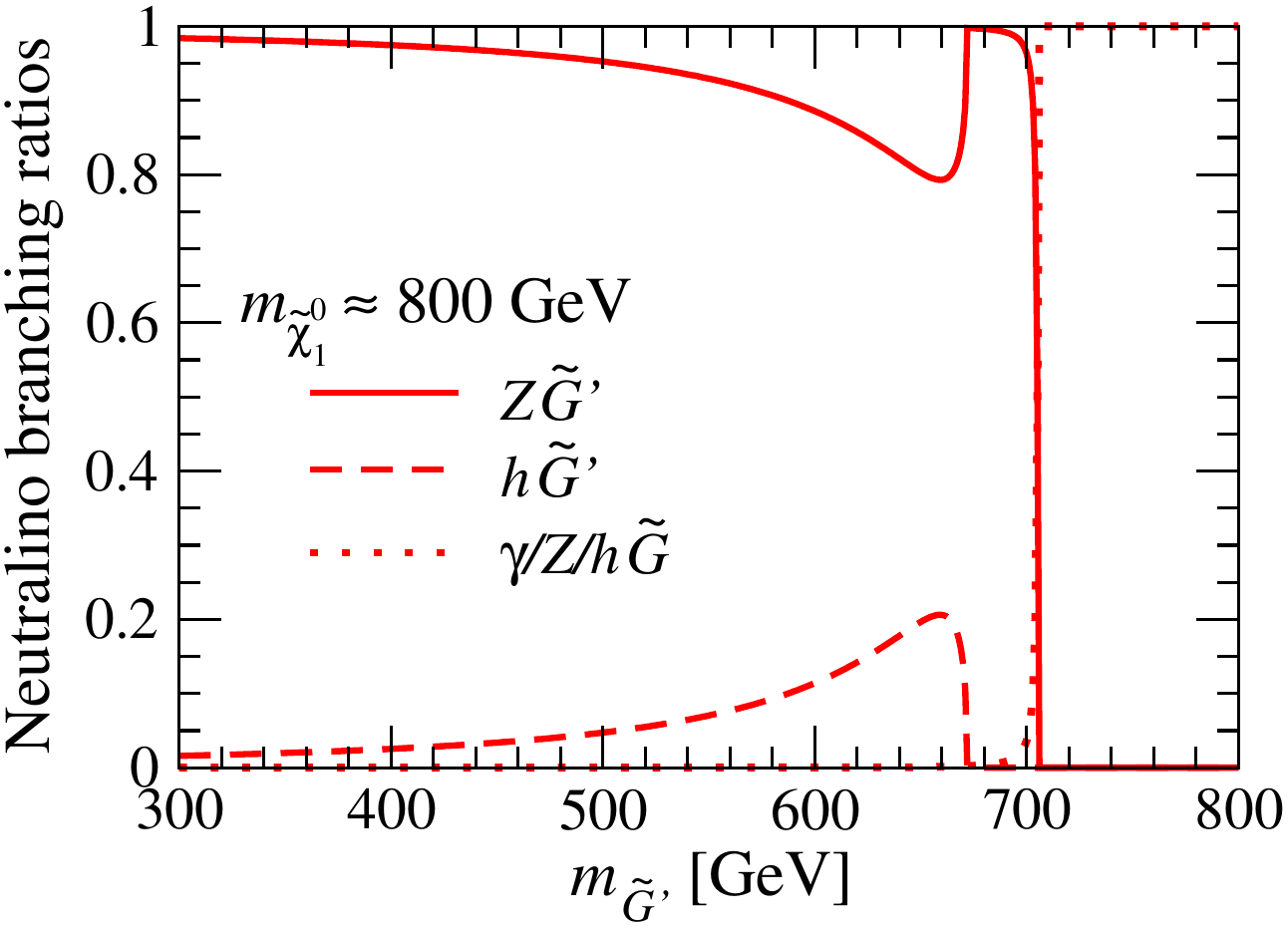}
\caption{\label{fig:branching}
 Branching ratios of the lightest neutralino as a function of the
 pseudo-goldstino mass for a representative benchmark point given 
 in~\ref{sec:neudecay}. 
 The $\tilde \chi^0_1 \to\gamma \pg$ decay is too small to see in the plot.
}
\end{figure}

In Fig.~\ref{fig:branching} we show the branching ratios of the lightest 
neutralino as a function of the pseudo-goldstino mass, evaluated on our
illustrative benchmark point described in detail in~\ref{sec:neudecay}. 
The decay pattern depends on the mass splitting 
$\Delta m_{\neuone-\pg} \equiv m_{\neuone} - m_{\pg}$
and on the possible kinematically allowed modes.
The branching ratio of $\neuone \to Z\pg$ is always greater than 80\% for
$\Delta m_{\neuone-\pg}>m_h$, and saturates at 100\% for
$m_Z<\Delta m_{\neuone-\pg}<m_h$. 
We note that the $\neuone \to\gamma\pg$ decay is negligible in our
parameter choice.   
In the regime of $\Delta m_{\neuone-\pg}<m_Z$, the decays into a true
goldstino plus a $\gamma$, $Z$ or $h$ become dominant due to the phase
space. 
In the following, we consider the region
\begin{align}
 \Delta m_{\neuone-\pg}>m_Z,
\label{dmn}
\end{align}
where the $\neuone \to Z\pg$ decay is dominant, as assumed in the
simplified model in Fig.~\ref{fig:spectra} and shown in Fig.~\ref{fig:branching}.

We have also verified that the total decay width in the region of
interest is always larger than $2 \times 10^{-12}$ GeV, implying that
the decays happen promptly in the detector.

Finally, the pseudo-goldstino will eventually decay into a goldstino
plus $\gamma$, $Z$ or $h$. 
However, one can verify that the pseudo-goldstino is enough long-lived
so that the decay happens outside the detector.
With the formulas listed in~\ref{sec:neudecay}, one can indeed compute
the pseudo-goldstino decay width and we find that for the benchmark
point it is around $10^{-22} -10^{-24}$~GeV, i.e. 
$\tau_{\pg}\lesssim 1$~sec, depending on the pseudo-goldstino mass. 
Even though we are not addressing cosmological issues in this paper, we 
observe that this decay is fast enough not to spoil the big bang
nucleosynthesis (BBN)~\cite{Kawasaki:2004qu}.

\section{Analyses}\label{sec:analyses}

\begin{figure}
\center
 \includegraphics[width=.8\columnwidth]{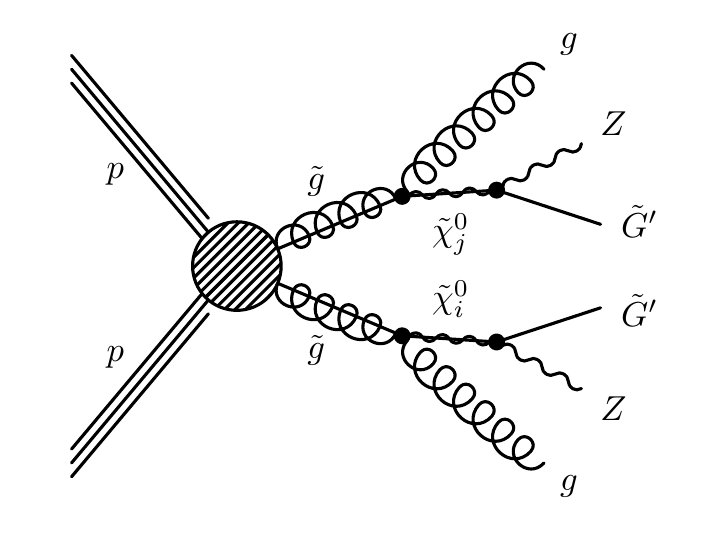}
\caption{\label{fig:diagram}
 The process at the LHC in our simplified goldstini model.}
\end{figure}

At the LHC, our simplified goldstini model can be probed by gluino
production.
Given the discussion on the decays in the previous section, the gluino
pair production leads to the process illustrated in
Fig.~\ref{fig:diagram}, where we assume the 100\% branching ratio at
each decay step as a good approximation.  
Depending on the decay of the $Z$ boson, the final state can be 
SFOS lepton pair + jets + $\MET$ and jets + $\MET$.

In this section, we investigate whether or not the event topology of our
model can successfully fit the ATLAS on-$Z$ excess without conflicting
with other searches. 
The most stringent constraints come from ATLAS multijet
search~\cite{Aad:2014wea} and CMS dilepton
search~\cite{Khachatryan:2015lwa} in the parameter region of our
interest~\cite{Ellwanger:2015hva,Allanach:2015xga,Lu:2015wwa}. 
The latter has signal regions which look at the same final state as in
the ATLAS on-$Z$ signal region and potentially quite constraining. 
We have implemented these analyses as well as the ATLAS on-$Z$ signal
region in {\tt Atom}~\cite{atom}.   
Some description and validation results of {\tt Atom} are given
in~\ref{sec:validation}.    

In order to fit the excess we scan the gluino mass having the neutralino
masses fixed at 
\begin{align}
 m_{\neu} = m_{\tilde g} - 200~{\rm GeV}.
\end{align}
We consider three cases featuring the pseudo-goldstino mass:
\begin{align}
 m_{\pg} = 
  \left\{ \begin{array}{ll}
    0 & ~~({\rm A}) \\
    m_{\neui} - 200~{\rm GeV} & ~~({\rm B}) \\     
    m_{\neui} - 100~{\rm GeV} & ~~({\rm C}) 
  \end{array} \right.
\label{eq:cases}
\end{align}
Case A is equivalent to the gauge mediation scenario with only one SUSY
breaking sector and a very light gravitino, while cases B and C have
compressed spectra.  

In order to assess the consistency between the model and data,
the ideal approach would be to carry out a global fit,
treating the excess and constraints in the same manner.
This requires the details of the systematic uncertainties and 
good understanding of the correlation among different signal regions.
Rather than taking this rigorous approach, in this exploratory paper 
we instead fit the model to the excess independently from the constraints 
and check the exclusion individually for the signal regions 
using the following prescription.

To see a goodness of the fit, we define the measure $R$ as
\begin{equation}
 R \equiv N_{\rm SUSY}/(N_{\rm obs} - N_{\rm SM}),
\end{equation}
for the ATLAS on-$Z$ signal region, where $N_{\rm SUSY}$ is the expected
SUSY events, $N_{\rm obs}$ is the number of observed events and 
$N_{\rm SM}$ is the expected SM events in the signal region. 
With this definition the best fit is given at $R=1$. 

For the other signal regions, labeled $i$, used as constraint, we instead
define
\begin{equation}
 R^i \equiv N^i_{\rm SUSY}/N^{{\rm UL},i}_{\rm BSM},
\end{equation}
where $N^{i}_{\rm SUSY}$ is the expected SUSY events and
$N_{\rm BSM}^{{\rm UL},i}$ is the 95\% CL$_s$ limit obtained from signal
region $i$. 
Having any $R_i$ greater than one indicates that the model is strongly
disfavoured. 

The $N^{(i)}_{\rm SUSY}$ can be expressed as
$\sigma_{\tilde g \tilde g} \cdot {\cal L} \cdot \epsilon_{(i)}$.
Here ${\cal L}$ is the integrated luminosity used in the analysis and 
$\sigma_{\tilde g \tilde g}$ is the production cross section of the
gluino pair, for which we use the values reported
in~\cite{susyXWG,Kramer:2012bx}.
To estimate the efficiency $\epsilon_{(i)}$ we use the following
simulation chain: 
first the signal events are generated using 
{\tt MadGraph5\_aMC@NLO}~\cite{mg5} and showered and hadronized by
{\tt Pythia6}~\cite{pythia6}.
The hadron level events are then passed to {\tt Atom}~\cite{atom} to
estimate the efficiency for each signal region taking the detector
effects into account. 
For our signal simulation, we extended the goldstini
model~\cite{Argurio:2011gu,Mawatari:2012ui} (building
on~\cite{Mawatari:2011jy}) to include the two-body gluino decay by using
{\tt FeynRules2}~\cite{Alloul:2013bka}.

\begin{figure*}
\center
 \includegraphics[width=.285\textwidth]{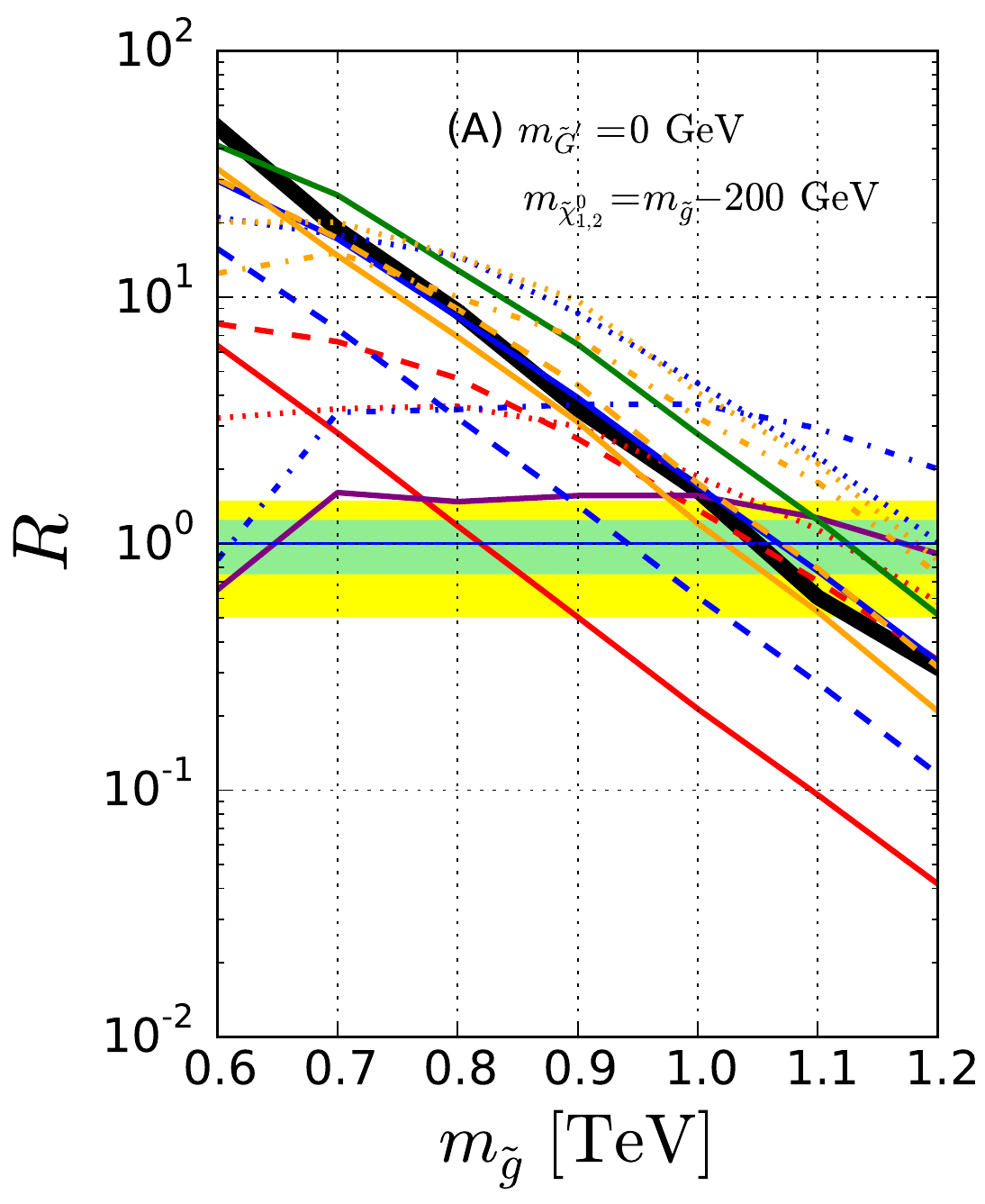}
 \includegraphics[width=.285\textwidth]{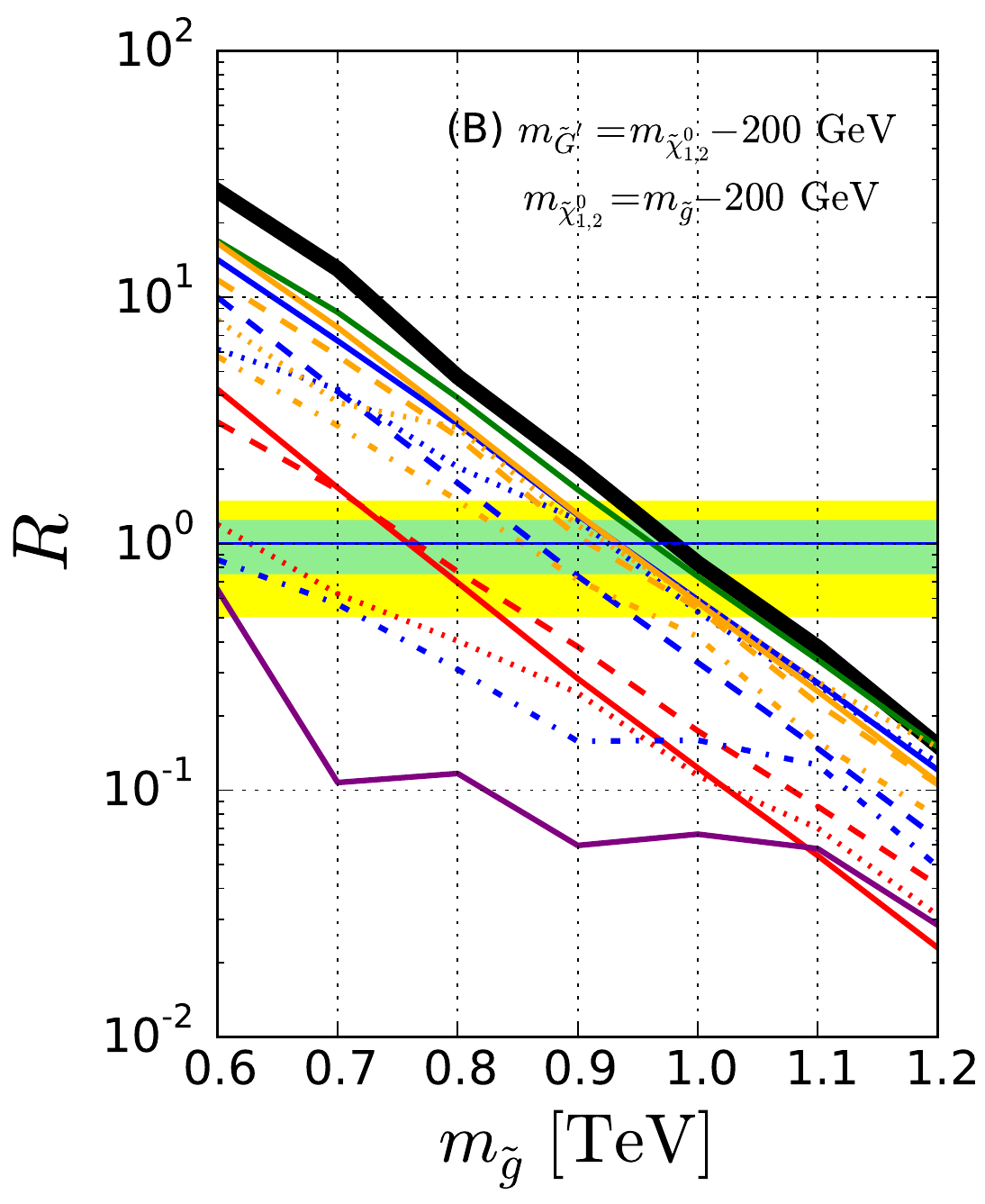}
 \includegraphics[width=.285\textwidth]{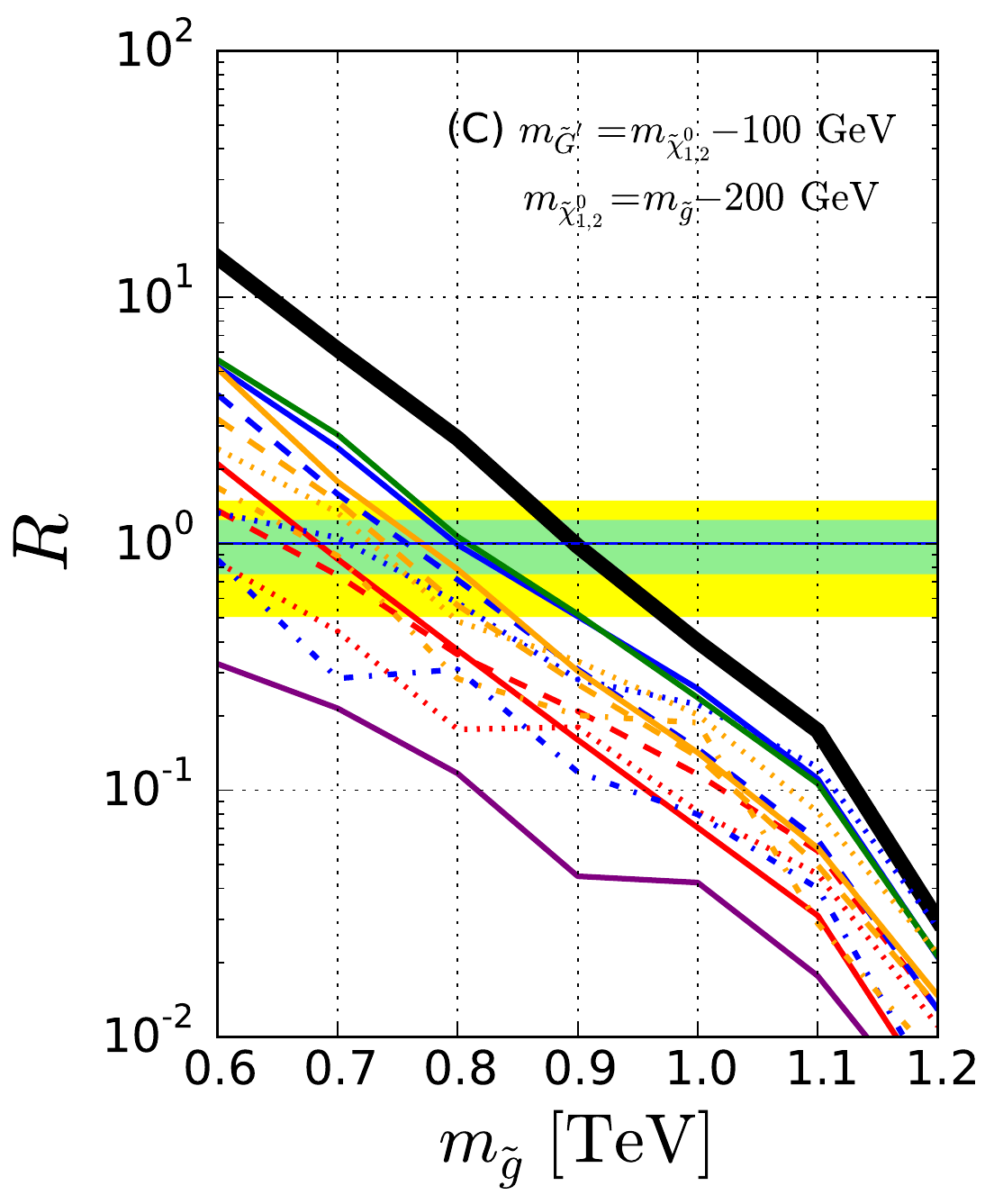}
 \raisebox{9mm}{\includegraphics[width=.07\textwidth]{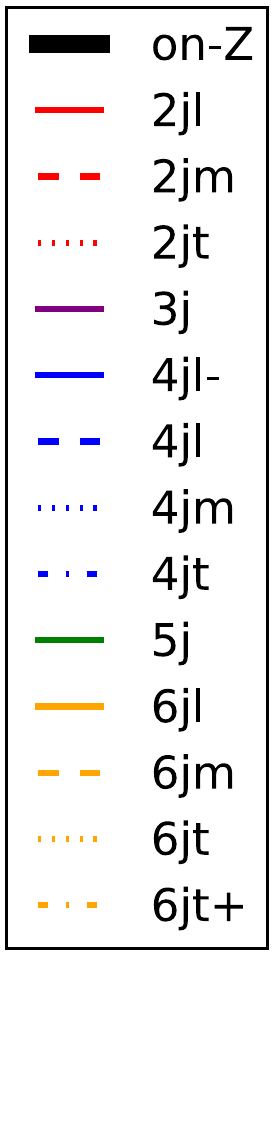}}
\caption{$R$ values for the fit (black solid) and constraints (others)
 for cases A (left), B (middle) and C (right) in Eq.~(\ref{eq:cases}). 
 The green and yellow bands correspond to the 1 and 2 $\sigma$ regions of
 the fit. 
 If the black curve is in the bands, the model provides a good fit
 for the ATLAS on-$Z$ excess~\cite{Aad:2015wqa}. 
 On the other hand, if there is any other curve above one, the model
 point is strongly disfavoured by the signal region corresponding to the
 curve in the ATLAS multijet + $\MET$ analyses~\cite{Aad:2014wea}.}
\label{fig:rvalue}
\end{figure*}

\begin{figure*}
\center
 \includegraphics[width=.285\textwidth]{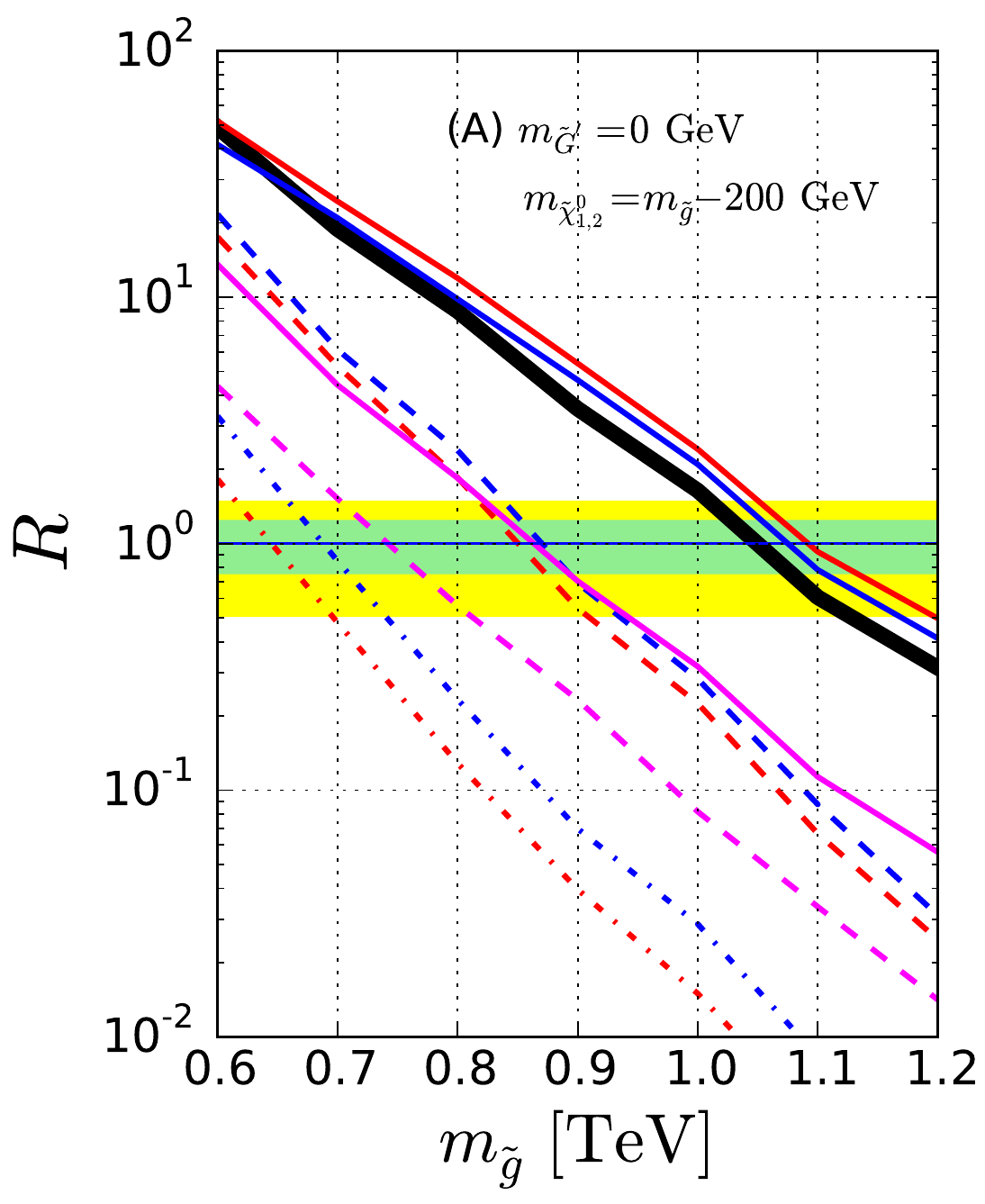}
 \includegraphics[width=.285\textwidth]{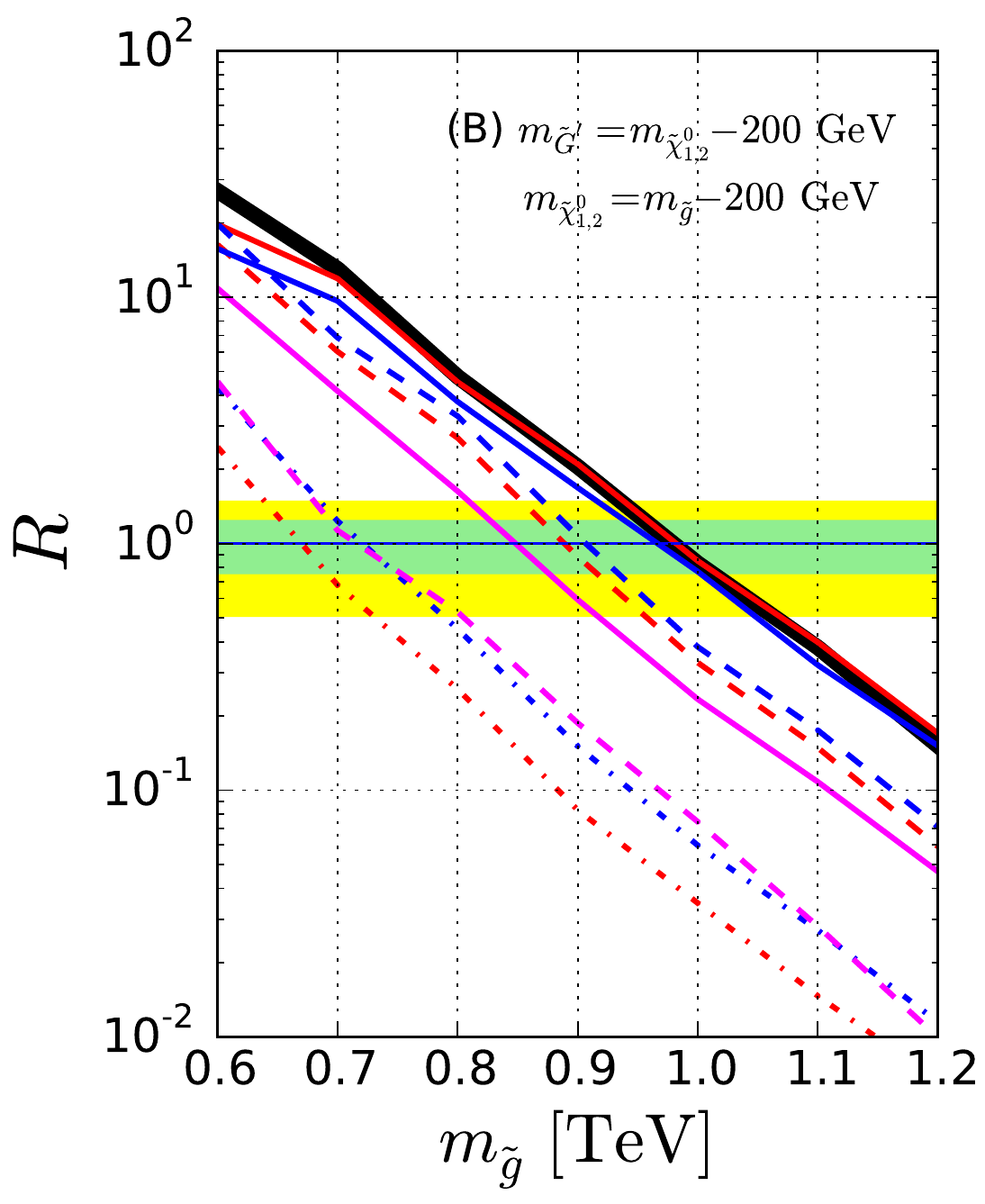}
 \includegraphics[width=.285\textwidth]{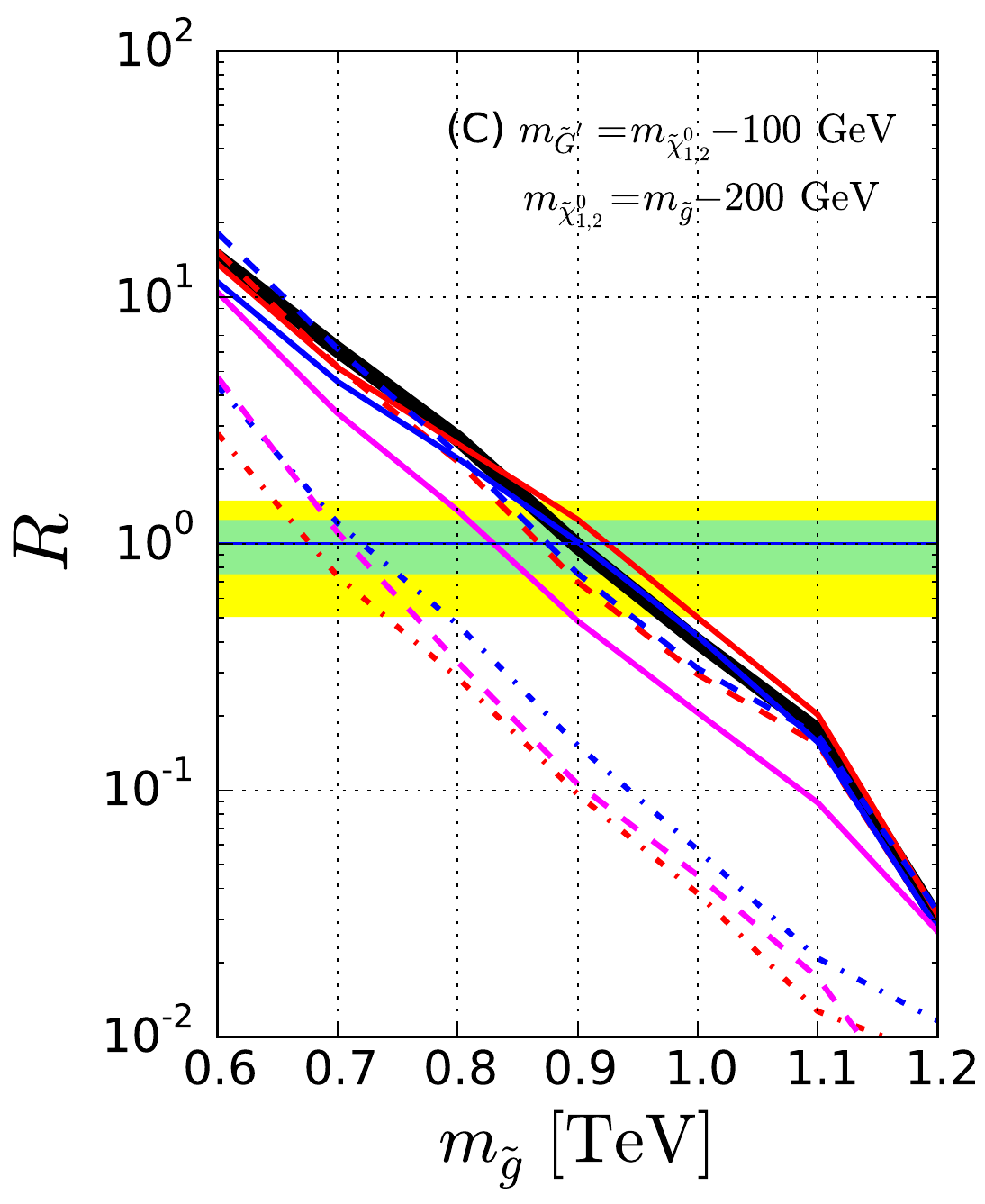}
 \raisebox{9mm}{\includegraphics[width=.09\textwidth]{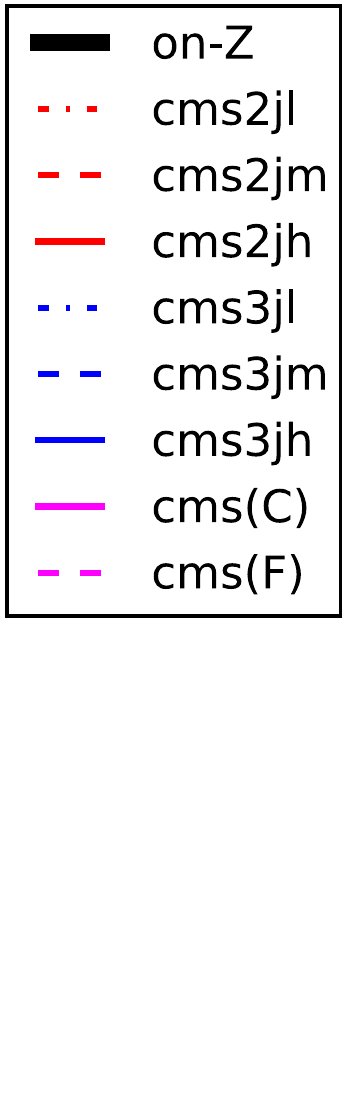}}
\caption{The same as in Fig.~\ref{fig:rvalue}, but for the constraints
 from the CMS on-$Z$ analyses~\cite{Khachatryan:2015lwa}.}
\label{fig:rvalue_cms}
\end{figure*}

The results of the three cases of the goldstini scenario in 
Eq.~\eqref{eq:cases} are presented in Figs.~\ref{fig:rvalue} and \ref{fig:rvalue_cms},
where the modes are confronted with the ATLAS multijet~\cite{Aad:2014wea} 
and the CMS dilepton~\cite{Khachatryan:2015lwa} searches, respectively.  
The fitting measure $R$ for the ATLAS on-$Z$ signal
region~\cite{Aad:2015wqa} is shown with the solid black curve, whereas
the constraints $R^{i}$ are shown with the other curves,
corresponding to the signal regions 
(2jl, 2jm, 2jt, 3j, 4jl-, 4jl, 4jm, 4jt, 5j, 6j, 6jm, 6jt, 6j+)
in the ATLAS multijet search in Fig.~\ref{fig:rvalue}
and 
(cms2jl, cms2jm, cms2jh, cms3jl, cms3jm, cms3jh, cms(C), cms(F))
in the CMS dilepton search in Fig.~\ref{fig:rvalue_cms}.%
\footnote{
The $n$j in the signal region name indicates it requires more than $n$
high $p_T$ jets. 
The letter ``l", ``m", ``t" for the ATLAS multijet search means
``loose", ``medium", ``tight", while ``l", ``m", ``h" for the CMS
dilepton search denotes ``low", ``medium", ``high".
The cms(C) and cms(F) represent the central and forward signal regions in which
the number of jets and $\MET$ are treated inclusively.
See~\cite{Aad:2014wea} and \cite{Khachatryan:2015lwa} for the exact definition.} 
The green (yellow) band around $R = 1$ represents the 1 (2) $\sigma$ region of the
fit for the ATLAS on-$Z$ excess. 

\begin{figure*}
\center
 \includegraphics[width=.415\textwidth]{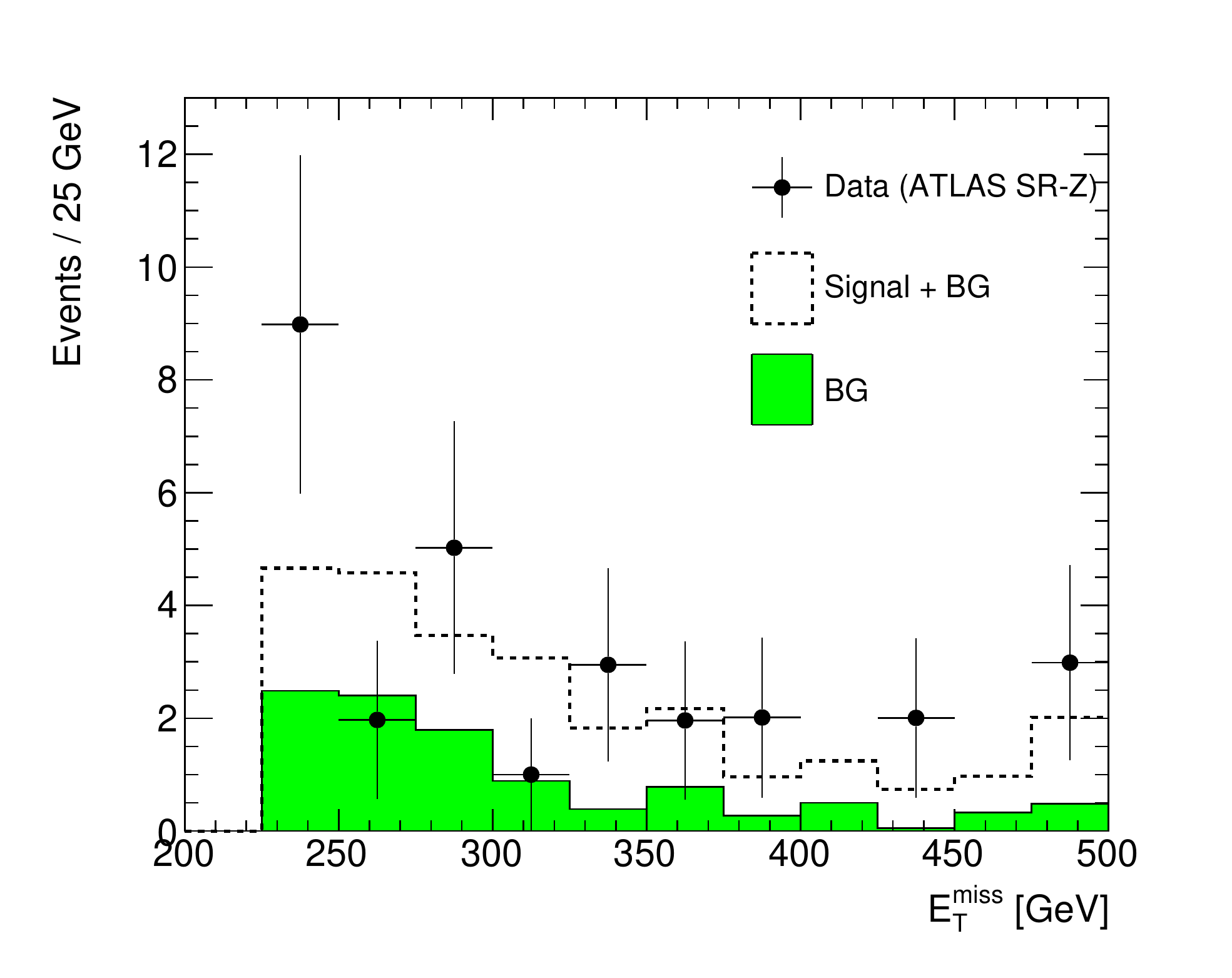}
 \includegraphics[width=.415\textwidth]{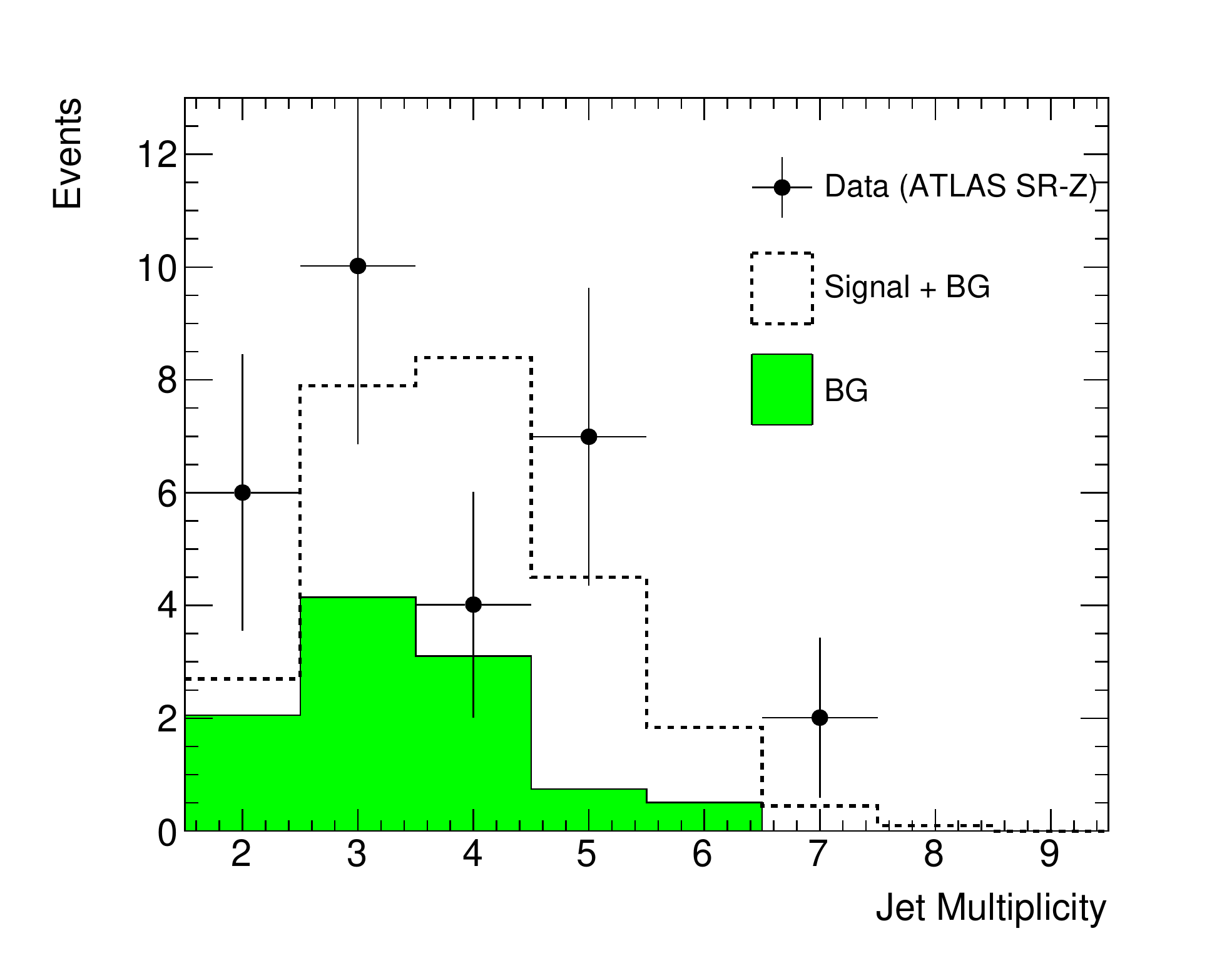}
\caption{Comparison between data and signal + background in the $\MET$
 (left) and jet multiplicity (right) distributions in the ATLAS on-$Z$
 signal region~\cite{Aad:2015wqa} at our best fit point:
 $(m_{\tilde{g}}, m_{\neu}, m_{\tilde{G'}}) = (1000, 800, 600)$~GeV.}
\label{fig:distributions}
\end{figure*}

\begin{figure*}
\center
 \includegraphics[width=.415\textwidth]{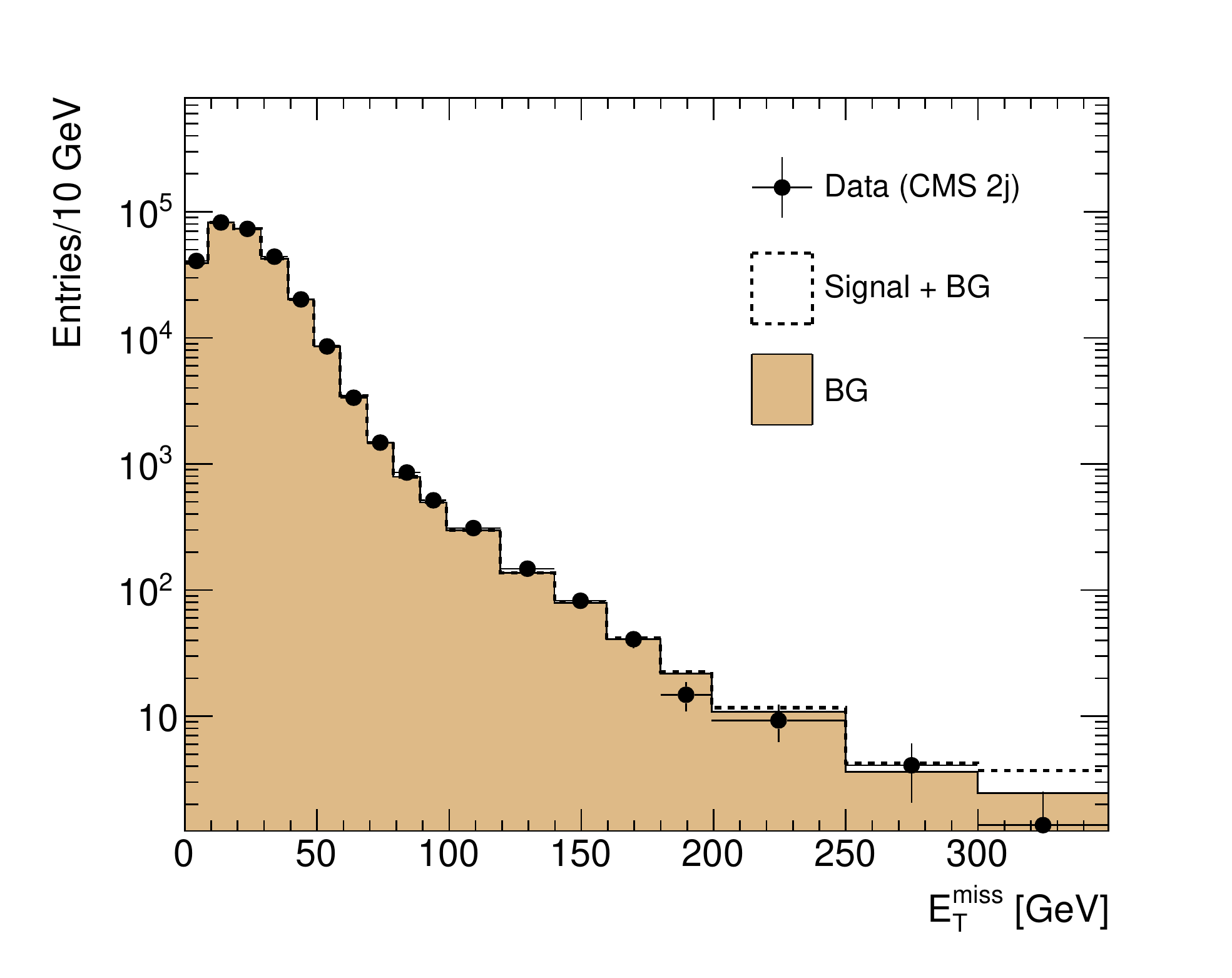}
 \includegraphics[width=.415\textwidth]{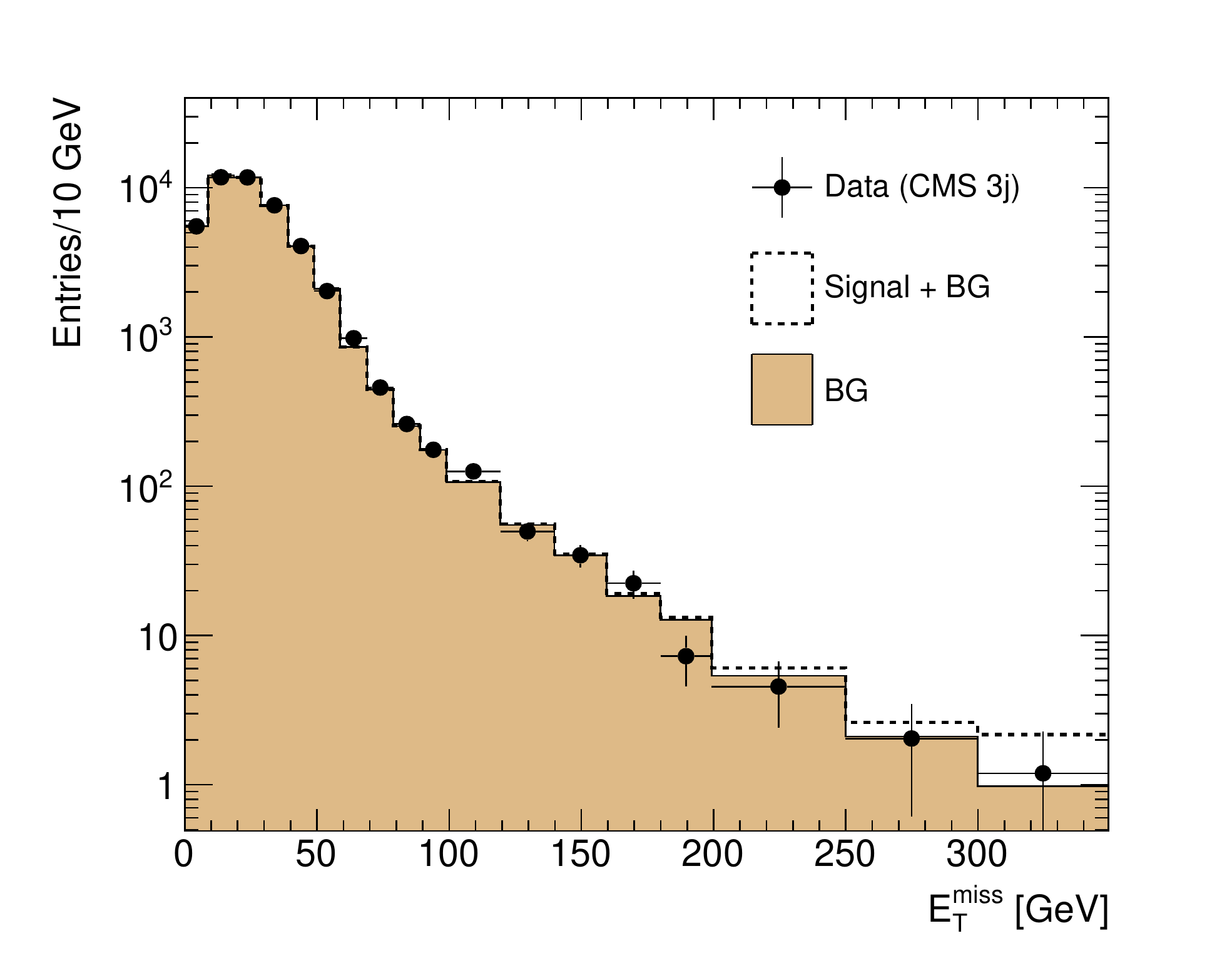}
\caption{
Comparison between data and signal + background in the $\MET$
 distribution in the CMS on-$Z$ signal region with $\ge2$ jets (left)
 and $\ge3$ jets (right)~\cite{Khachatryan:2015lwa} at our best fit
 point: 
 $(m_{\tilde{g}}, m_{\neu}, m_{\tilde{G'}}) = (1000, 800, 600)$~GeV.
}
\label{fig:distributions_cms}
\end{figure*}
 
One can see that the entire region of case A, i.e. the massless
goldstino case, is excluded by the ATLAS multijet search.
The CMS dilepton search also excludes the region with $m_{\tilde g} < 1.1$~TeV.
These strong limits can be attributed to the large mass gap between the gluino and
the pseudo-goldstino, because of which the jets and leptons from the gluino decays
tend to be hard, helping to increase the efficiency of the ATLAS multijet search
as well as the CMS dilepton search.

Moving to case B and C, one can see that the constraints are generally more relaxed
than in case A, since these cases have milder mass hierarchy with the massive pseudo-goldstino.
The ATLAS multijet search now excludes the gluino mass up to 970 (810)~GeV for case B (C).
Here, the best fit of the ATLAS on-$Z$ excess is given at
$m_{\tilde g} = 980$ (900)~GeV for case B (C), which is just outside of 
the ATLAS multijet exclusion limit.

The CMS dilepton search, on the other hand, still provides tight
constraints, especially the cms2jh signal region, where $\MET>300$~GeV
is required, very similar to the $\MET>225$~GeV cut in the ATLAS on-$Z$
analysis. 
The difference between the two analyses is that the ATLAS search
additionally imposes $H_T > 600$~GeV, where $H_T$ is the scalar sum of 
the $p_T$ of jets and leptons.  
For case B, the best fit point ($m_{\tilde g} = 980$~GeV) is just on the
exclusion boundary,
while for case C the best fit point ($m_{\tilde g} = 900$~GeV) is just excluded by the cms2jh signal region.
For case B, at $m_{\tilde g} = 1$~TeV, the fit for the ATLAS on-$Z$ excess is still within the 1~$\sigma$ band, and the point is still outside of the 95\% ${\rm CL}_s$ exclusion. 
We therefore choose our best fit benchmark point as
\begin{align}
 (m_{\tilde{g}}, m_{\neu}, m_{\tilde{G'}})
 = (1000, 800, 600)~{\rm GeV}
\label{bestpoint}
\end{align}
for the following analysis.
Even for case C, the tension between the data and the prediction observed 
in the ATLAS on-$Z$ signal region
can be ameliorated to the 2~$\sigma$ level with $m_{\tilde g} = 950$~GeV,
which is 
outside the 95\% ${\rm CL}_s$ exclusion region.

In Fig.~\ref{fig:distributions} we compare the data with the signal +
background in the $\MET$ (left) and jet multiplicity (right)
distributions in the ATLAS on-$Z$ signal region at our best fit point 
in~\eqref{bestpoint}.
Here we took the data and the SM background from Figs.~6 and 7 in the
ATLAS paper~\cite{Aad:2015wqa} and combined $ee$ and $\mu\mu$
channels. 
The data in the $\MET$ distribution has a preference of low $\MET$,
peaking around 240~GeV and 
roughly
falling down up to around 500~GeV. 
The distribution is well fitted with the signal + background at our best
fit point because the massive nature of the pseudo-goldstino reduces the
$\MET$ in the event.  
In the jet multiplicity distribution the data prefers $2-5$ jets and
disfavors the region with $\geq 6$ jets.   
The distribution of the signal + background at our best fit point peaks
around $3-4$ and gives a good fit to the data. 
This is an advantage of the radiative decay 
$\tilde g \to g \tilde \chi_{1,2}^0$ compared to the three-body
$\tilde g \to q \bar q \tilde \chi_{1,2}^0$ decay because the number of
jets is reduced typically by two.  

As a reference, in Fig.~\ref{fig:distributions_cms} we also compare the
signal with the CMS data in the
on-$Z$ signal region with $N_{\rm jets}\ge2$ and $\ge3$ at our best fit
point in~\eqref{bestpoint}. 
Here we took the data and the SM background from Fig.~7 in the
CMS paper~\cite{Khachatryan:2015lwa}. 
As already seen in Fig.~\ref{fig:rvalue_cms}, the most stringent
constraint comes from the high $\MET$ region, i.e. cms2jh and cms3jh,
where the $\MET>300$~GeV bin is considered.

\section{Summary}\label{sec:summary}

Even though LHC at 8 TeV has not discovered a new physics signal, the
Run-I data contain a few small excesses that deserve a thorough
investigation. 
Recently the ATLAS Collaboration has reported a 3.0 $\sigma$ excess in
dilepton + jets + $\MET$, with the dilepton reconstructing the $Z$-boson mass.

In this paper we have proposed an explanation of such excess in a SUSY
model of gauge mediation with two SUSY breaking sectors, 
presenting in the SUSY spectrum an extra neutral fermion besides the
MSSM neutralinos, that is the pseudo-goldstino. 
Our simplified model consists in a gluino, a pair of Higgsino-like
neutralinos, and a pseudo-goldstino. 
We showed that our goldstini model can explain the ATLAS $Z$-peaked
excess without conflicting with the constraints from multijet + $\MET$ as
well as from the CMS analysis for the same final state.
The compressed spectrum such as $m_{\tilde g}\sim 1000$~GeV,
$m_{\tilde\chi^0_{1,2}}\sim 800$~GeV and $m_{\tilde G'}\sim 600$~GeV
gives a very good fit to the data not only for the rate but also for the
distributions. 

If this is indeed the origin of the excess, the 13-TeV LHC with both the
ATLAS and CMS Collaborations will soon confirm the excess with much
greater significance.
We await the verdict from the LHC Run-II.

\section*{Acknowledgments}

We would like to thank Gabriele Ferretti, Diego Redigolo, and Pantelis
Tziveloglou for helpful discussions, and
Satoshi Shirai for valuable comments on the cosmological issue. 
We are particularly grateful to Yevgeny Kats for useful suggestion on
the CMS constraints.

S.P.L. is supported by JSPS Research Fellowships for Young Scientists
and the Program for Leading Graduate Schools, MEXT, Japan. 
A. M. is a Pegasus FWO postdoctoral Fellowship. 
K.S. is supported in part by
the London Centre for Terauniverse Studies (LCTS), using funding from
the European Research Council 
via the Advanced Investigator Grant 267352.
M.V is an aspirant FWO-Vlaanderen.
K.M., A. M. and M.V. are also supported in part by the Strategic
Research Program High Energy Physics and the Research Council of the
Vrije Universiteit Brussel.

\appendix
\section{Neutralino decay}\label{sec:neudecay}

In this appendix, we review the true goldstino and the pseudo-goldstino
couplings with the MSSM states, and derive the formulas for the relevant
decay widths; see also
Refs.~\cite{Cheung:2010mc,Thaler:2011me,Argurio:2011gu,Liu:2013sx,Ferretti:2013wya}.

In the two-sector goldstini model, the lagrangian relevant to the
neutralino decay, in the gauge eigenstate 
$(\tilde\psi_i=\{\tilde B,\tilde W, \tilde H_d,\tilde H_u\})$,
is%
\footnote{We follow the conventions of~\cite{Martin:1997ns}.}
\begin{align}
\mathcal{L} \supset &-\frac{1}{2} M_{ij}^{\tilde\psi} \tilde \psi_i \tilde \psi_j
 -\rho_{ai} \tilde G_a \chi_i -\frac{1}{2} M_{ab}^{\tg} \tilde G_a \tilde G_b -\frac{1}{2} Y_{ij} \tilde \psi_i \tilde \psi_j h_0  \nonumber \\
&+ \tau_{ai} \tilde G_a \tilde \psi_i h_0+G_{ij} \tilde \psi_i^{\dagger} \bar \sigma^{\mu} \tilde \psi_j Z_{\mu}+  i L_{ia} \tilde \psi_i   \sigma^{\mu} \bar \sigma^{\nu} \tilde G_a Z_{\mu\nu},
\label{gauge_lag}
\end{align}
where $M_{ij}^{\tilde\psi}$ is the standard $4\times 4$ neutralino mass matrix. 
The other parameters are
\begin{align}
&\rho_{a}=-\frac{1}{\sqrt{2}f_a}\left(
\begin{array}{c}
 M_{B(a)} \langle D_Y \rangle \\
 M_{W(a)} \langle D_{T^3}\rangle \\
 \sqrt{2}v \big( m_{H_d(a)}^{2} c_{\beta}  - B_{(a)} s_{\beta} \big)\\
 \sqrt{2}v \big( m_{H_u(a)}^{2} s_{\beta} - B_{(a)} c_{\beta} \big)
\end{array}
\right),\\
&Y = \frac{1}{2}
\left( 
\begin{array} {cccc}
 0 & 0 & -g_1 c_{\beta}  & g_1 s_{\beta}  \\
 0 & 0 & g_2 c_{\beta}  & -g_2 s_{\beta}  \\
 -g_1 c_{\beta}  & g_2 c_{\beta}  & 0 & 0 \\
 g_1 s_{\beta}  & -g_2 s_{\beta} & 0 & 0 \\
\end{array}
\right),\\
&\tau_{a}=\frac{1}{\sqrt{2}f_a}
\left(
\begin{array}{c}
  m_Z M_{B(a)}s_W c_{ 2 \beta}   \\
 -m_Z M_{W(a)}c_W c_{2 \beta}  \\
  m_{H_d(a)}^{2} c_{\beta}  - B_{(a)} s_{\beta} \\
  m_{H_u(a)}^{2}  s_{\beta}  - B_{(a)} c_{\beta} \\
\end{array}
\right),\\
&G=\frac{g_2}{2 c_W} 
\left(
\begin{array}{cccc}
0 & 0 & 0 & 0 \\
0 & 0 & 0 & 0 \\
0 & 0 & 1 & 0 \\
0 & 0 & 0 & -1 \\
\end{array}
\right),\\
&L_{a}=\frac{1}{2 \sqrt{2} f_a}
\left(
\begin{array}{c}
 -M_{B(a)} s_W \\
  M_{W(a)} c_W \\
  0 \\
  0
\end{array}
\right),
\end{align}
where we denote with $M_{B/W(a)}$, $m^2_{H_{d/u}(a)}$ and $B_{(a)}$ the
contribution to the soft term from sector $a$, 
and consider the decoupling limit for the Higgses. 
For convenience it is also useful to define tilded soft masses, e.g.
\begin{align}
\tilde M_B =-\frac{f_2}{f_1} M_{B(1)}
 +\frac{f_1}{f_2} M_{B(2)},
\end{align}
while the non-tilded are the physical ones, e.g. 
\begin{align}
 M_B=M_{B(1)}+M_{B(2)}.
\end{align}
The mass matrix $M^{\tg}$ is 
\begin{align}
 M^{\tg}=
\left(
\begin{array}{cc}
 -\frac{f_2}{f_1} \mathcal{M}_{12} & \mathcal{M}_{12} \\
 \mathcal{M}_{12} & -\frac{f_1}{f_2} \mathcal{M}_{12} \\
\end{array}
\right),
\end{align}
and it is such that it has one zero eigenvalues along the true goldstino
eigenstate in Eq.~\eqref{gld_pgld}. 
The entry $\mathcal{M}_{12}$ is a model dependent quantity that has been
computed in~\cite{Argurio:2011hs}. 
The other eigenvalue of $M^{\tg}$ is the pseudo-goldstino mass
$m_{\pg}$. 

Given a set of soft terms and the relative contributions from the two
sectors, one should diagonalize the lagrangian and write the resulting
couplings in the mass eigenbasis.  
This ends up in the lagrangian quoted in Eq.~\eqref{pgld_couplings},
whose couplings can be computed for a given set of soft terms. 
Note that, even though suppressed by $1/f_a$, there will be some mixing
between the neutralinos and the goldstini. 
This is the reason why we have also written in~\eqref{gauge_lag} the
terms containing only the neutralino gauge eigenstates. 
For instance, once rotated into the mass eigenbasis, these interactions
generate the effective couplings $\tilde y_{Z_L}$ in 
Eq.~\eqref{pgld_couplings}.

Although we study a rather heavy pseudo-goldstino case in this paper,   
we consider the $m_{\pg}=0$ limit for a while to get an insight on the
feature of the pseudo-goldstino couplings by the analytic expressions.
At leading order in $m_Z/m_{\tilde \chi}$ in the neutralino and
pseudo-goldstino mixing and neglecting the pseudo-goldstino mass, the
effective couplings in the pseudo-goldstino
lagrangian~\eqref{pgld_couplings} are  
\begin{align}
 \tilde y^i_{\gamma} &= m_{\neui}  ( K_B N_{i1} c_W+K_W N_{i2} s_W) \nn\\
  &\quad +m_Z (N_{i4}^* s_{\beta}-N_{i3}^* c_{\beta}) s_W c_W (K_B-K_W), \\
 \tilde y^i_{Z_T} &= m_{\neui}  ( -K_B N_{i1} s_W+K_W N_{i2} c_W) \nn\\
  &\quad +m_Z (N_{i4}^* s_{\beta}-N_{i3}^* c_{\beta}) (s_W^2 K_B+c_W^2 K_W), \\
\tilde y^i_{Z_L} &= -m_Z K_{\mu} ( - N_{i1} s_W+N_{i2} c_W) \nn\\
  &\quad -m_{\neui}  (K_u N_{i4}^* s_{\beta}-K_d N_{i3}^* c_{\beta}), \\
 \tilde y^i_{h} &=-m_Z \cos 2 \beta ( -K_B M_B N_{i1}^* s_W+K_W M_W N_{i2}^* c_W)  \nonumber \\
&  \quad -\mu^2 (K_d N_{i3}^* c_{\beta} + K_u N_{i4}^* s_{\beta} ),
\end{align}
where the neutralino rotation matrix $N_{ij}$ is defined as
 in~\cite{Martin:1997ns}.
The $K$ factors read
\begin{align}
 K_B &= \frac{\tilde M_B}{M_B}, \quad 
 K_W=\frac{\tilde M_W}{M_W}, \quad
 K_{\mu} =c_{\beta}^2 K_d + s_{\beta}^2 K_u, \nn\\
 K_{d} &=-\frac{1}{\mu^2} \big(\tilde m_{H_d}^2-\tilde B \tan \beta
 \nn\\
 &\hspace*{14mm} +\frac{m_Z^2}{2} (s_W^2 K_B+c_W^2 K_W) \cos 2 \beta \big), \nonumber \\
 K_{u} &=-\frac{1}{\mu^2} \big(\tilde m_{H_u}^2-\tilde B \cot \beta
 \nn\\
 &\hspace*{14mm}-\frac{m_Z^2}{2} (s_W^2 K_B+c_W^2 K_W) \cos 2 \beta \big),
\label{kfactor}
\end{align}
where we keep the terms of order $m_Z^2$ in the last two $K$ factors in
order to show the goldstino limit explicitly. 
The goldstino 
lagrangian~\cite{Martin:1997ns,Ambrosanio:1996jn,Komargodski:2009rz,Luo:2010he,Antoniadis:2010hs}
is recovered by converting all the tilded soft terms to the un-tilded
ones, and using the electroweak symmetry breaking (EWSB) conditions
\begin{align}
 0 &= m_{H_u}^2+\mu^2-B \cot \beta -\frac{1}{2} m_Z^2 \cos 2\beta, \\
 0 &= m_{H_d}^2+\mu^2-B \tan \beta +\frac{1}{2} m_Z^2 \cos 2\beta,
\end{align}
resulting in all the $K$ factors in Eq.~\eqref{kfactor} to be unity.
From the above analytic couplings, for instance, one can observe that
for a mostly Higgsino-like neutralino with $K_B=K_W$ the
pseudo-goldstino coupling to the photon is suppressed.  
We also checked that the above expressions agree with the numerics if
the pseudo-goldstino mass is sufficiently smaller than the neutralino
masses.  
For the large pseudo-goldstino mass, on the other hand, there can be
deviations from these formulas once we rotate in the mass eigenbasis. 
Hence, in order to compute correctly the effective couplings in
Eq.~\eqref{pgld_couplings} for generic pseudo-goldstino masses, one is
instructed to diagonalize numerically the lagrangian~\eqref{gauge_lag},
for a given set of soft terms, and write it in the mass eigenbasis.  

Once the effective couplings in Eq.~\eqref{pgld_couplings} are
evaluated, the neutralino decay widths can be computed 
as~\cite{Thaler:2011me,Argurio:2011gu} 
\begin{align}
\label{decaya}
 \Gamma_{\neui\to\gamma\pg} &= 
  \frac{(\tilde y^i_{\gamma})^2 m_{\neui}^3}{16 \pi F^2} 
  \bigg(1-\frac{m_{\pg}^2}{m_{\neui}^2} \bigg)^3, \\
\label{decayz}
 \Gamma_{\neui\to Z\pg}&= \frac{\beta_Z m_{\neui}}{32 \pi F^2} 
   \bigg[\bigg(1-\frac{m_{\pg}}{m_{\neui}}\bigg)^2-\frac{m_Z^2}{m_{\neui}^2}\bigg]  \nn\\
  &\quad\times\big[ (\tilde y^i_{Z_T})^2 (2 (m_{\neui}+m_{\pg})^2+m_Z^2)  \nonumber \\
  &\qquad +(\tilde y^i_{Z_L})^2 ((m_{\neui}+m_{\pg})^2+2m_Z^2) \nn\\
  &\qquad +6 \tilde y^i_{Z_T} \tilde y^i_{Z_L} m_Z(m_{\neui}+m_{\pg})
\big], \\
\label{decayh}
 \Gamma_{\neui\to h\pg} &=\frac{ \beta_h (\tilde y^i_{h})^2  m_{\neui}}{32\pi F^2}
 \bigg(1+3 \frac{m_{\pg}^2}{m_{\neui}^2}-\frac{m_h^2}{m_{\neui}^2}\bigg),
\end{align}
where $\beta_{Z/h}=\bar\beta(m_{\pg}^2/m_{\neui}^2,
        m_{Z/h}^2/m_{\neui}^2)$
with
$\bar\beta(a,b)=(1+a^2+b^2-2 a  - 2 b  - 2 a b)^{1/2}$.
The standard decays into the massless goldstino are obtained from this
expression by sending the tilded quantities to the non-tilded quantities
and putting $m_{\pg}\to0$.

In this paper we are interested in configurations where the two lightest
neutralinos are mostly Higgsinos and the decay of $\neuone$ is predominantly into a
massive pseudo-goldstino and a $Z$ boson. 
To find if this scenario is compatible with the lagrangian just
explained, we numerically explored the parameters of the model (i.e. the
soft terms) looking for a representative benchmark point satisfying
these requirements.  
In Fig.~\ref{fig:branching} we report the branching ratios for a
configuration with $\mu=804$~GeV, $\sqrt{B}=800$~GeV, $M_B=M_W=1.5$~TeV
and $\tan\beta=10$.   
The total soft masses for the Higgses are extracted by solving the EWSB
conditions. 
The two SUSY breaking scales are chosen as
$\sqrt{f_1}=1.5\times10^6$~GeV and $\sqrt{f_2}=5 \times10^4$~GeV. 
The gaugino masses and the Higgs soft masses are distributed in the two
sectors as  
$M_{B/W(1)}/M_{B/W(2)}=\tan^2\theta$, 
$m_{H_{d/u}(1)}^2/m_{H_{d/u}(2)}^2=\cot^2\theta$.
The angle $\theta$ is taken to be $\tan \theta=0.2$.
Finally the $B$ terms are chosen as $B_{(a)}=f_a/(f_1+f_2) B$.
With these values we obtain a scenario where the two lightest neutralinos
are mostly Higgsinos, with masses at $(797,805)$~GeV. 
The lightest neutralino decay is predominantly into a pseudo-goldstino plus a $Z$ boson for
different ranges of pseudo-goldstino mass, as we show in Fig. \ref{fig:branching}. 
The pseudo-goldstino mass is varied by changing $\mathcal{M}_{12}$ 
consistently with the perturbative definition in~\cite{Argurio:2011hs}.
The decay width of the neutralinos cited in the text are computed with
the decay formulas~\eqref{decaya}, \eqref{decayz} and \eqref{decayh}. 
Moreover, the decay of the pseudo-goldstino can be computed using the
same formula quoted above, where the coupling between the
pseudo-goldstino and the goldstino are extracted numerically from the
original lagrangian once we switch to the mass eigenbasis.

\section{Analysis implementation \label{sec:validation}}

For this paper, we implemented several ATLAS and CMS analyses in the
{\tt Atom} framework~\cite{atom}.
{\tt Atom} is a program based on {\tt Rivet}~\cite{rivet} and maps the
truth level particles into the reconstructed objects such as isolated
electrons and $b$-jets according to the detector performances reported
by ATLAS and CMS.    
The validation and some application of the code can be found 
in~\cite{Papucci:2011wy,Papucci:2014rja,Kim:2014eva,Grothaus:2015yha,Jacques:2015zha}.

In Table~\ref{tbl:val} we show some of the validation results as an example.
The numbers in the second column represent the expected signal events
for each step of the cut used in the 5j signal region (SR) reported by
the ATLAS multijet analysis~\cite{Aad:2014wea},
based on the
$\tilde q_L \to q \tilde \chi_1^\pm \to qW^\pm  \tilde \chi_1^0$
topology with 
$(m_{\tilde q}, m_{\tilde \chi_1^\pm}, m_{\tilde \chi_1^0}) = (665,465,265)$ GeV.  
The right column shows the ratios between the {\tt Atom} and ATLAS
results. 
One can see that these ratios are close to one within $\sim 20$\%
accuracy indicating a good agreement between the {\tt Atom} and ATLAS
simulations. 

\begin{table}[h] 
\center
\begin{tabular}{|l|r|r|}
\hline
5j SR cuts& $N_{\rm SUSY}^{\rm Exp}$ &{\tt Atom}/Exp   
\\
\hline
$\MET > 160,\, p_T^{j_{1(2)}} > 130(60)\,$GeV & 317.3  &1.17
\\
$p_T^{j_3} > 60\,$GeV & 306.2  &1.12
\\
$p_T^{j_4} > 60\,$GeV & 247.6 &1.04
\\
$p_T^{j_5} > 60\,$GeV & 141.8 & 1.00 
\\
$\Delta\phi (j_{1,2,3},\MET) > 0.4 $& 118.6& 1.01
\\ 
$\Delta\phi (j_{i>3}>40\,{\rm GeV},\MET) > 0.2 $& 103.1& 1.01
\\
$\MET/m_{\rm eff}(N_j)>  0.2$ &85.6 &1.04
\\
$m_{\rm eff}({\rm incl.})> 1200\,$GeV &20.5 &1.18
\\
\hline
\end{tabular} 
\caption{ 
 ``5j'' SR validation table for the implementation of the ATLAS multijet
 analysis in {\tt Atom}.
 The decay chain for validation in consideration is 
 ${\tilde q}_L \to q\tilde{\chi}_1^{\pm} \to qW^{\pm}\tilde{\chi}_1^0$. 
\label{tbl:val}
}
\end{table}

\bibliography{bibgoldstini}

\end{document}